\documentclass[
    ,final            
  ]
  {aipproc}

\layoutstyle{8x11double}

\begin{document}

\title{Open Cavity Solutions to the rf in Magnetic Field Problem}


\author{Robert B.~Palmer, J. Scott~Berg, Richard C.~Fernow, Juan C.~Gallardo, Harold G.~Kirk}{
  address={Brookhaven National Laboratory\\
Upton, New York 11973, USA}
}

\begin{abstract} 
It has been observed~\cite{break} that breakdown in an 805~MHz pill-box cavity occurs at much lower gradients as an external axial magnetic field is increased. This effect was not observed with on open iris cavity. It is proposed that this effect depends on the relative angles of the magnetic and maximum electric fields: parallel in the pill-box case; at an angle in the open iris case. If so, using an open iris structure with solenoid coils in the irises should perform even better. A lattice, using this principle, is presented, for use in 6D cooling for a Muon Collider. Experimental layouts to test this principle are proposed.


\end{abstract}

\maketitle


\section{Introduction}

\begin{figure}[hbt!]
\includegraphics*[width=75mm]{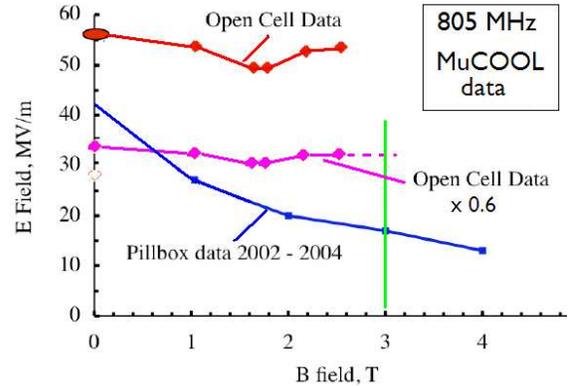}
\caption{(Color) Observed breakdown gradients in: a pill-box cavity (blue); an open cavity (red); the open cavity gradient multiplied by 60\% to give an estimated average acceleration (magenta).}
\label{break}
\end{figure}

An 805~MHz pill-box cavity has been shown~\cite{break} to have rapidly falling maximum gradient as the axial magnetic field is increased (see figure~\ref{break}).
If this relative drop is present at 201 MHz, it will preclude operation of the current ISS~\cite{iss} Neutrino Factory and RFOFO 6D Muon Collider cooling schemes. The earlier Lab G  tests with  a multi-cell open cavity design did not show~\cite{break} such a fall off (see figure~\ref{break}).
The average acceleration in an open cell structure is less ($\approx$ 60\%) of the maximum surface field, but even with this penalty, for magnetic fields above $\approx$ 1~T, the average acceleration appears to be higher in an open cell structure than for the pill-box.

A possible explanation of these observations depends on the angles between the maximum electric fields and the magnetic fields where they occur: parallel in the pill-box case; at an angle in the open iris case.  If this is the explanation, then even better performance would be expected if the solenoid coils were placed in the open cavity irises in such a way that the magnetic fields are almost perpendicular to the maximum electric gradients. This situation has been referred to as \emph{magnetic insulation}~\cite{mag-ins1},~\cite{mag-ins2}.

As an example, we will discuss a modification of the RFOFO lattice proposed as the first 6D cooling stage  in ref.~\cite{pac}. We will then discuss a possible R\&D program to study this possible solution.

\begin{figure}[hbt!]
\centering
\includegraphics*[width=75mm]{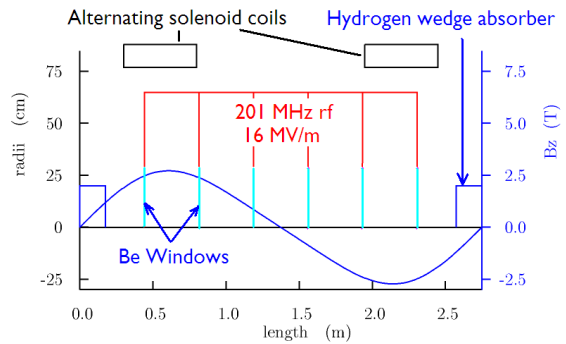}
\caption{(Color) Schematic of one RFOFO cell used in Ref.~\cite{pac}.}
\label{rfofo-cell}
\end{figure}

\section{Example of First RFOFO 6D Cooling lattices}
The \texttt{RFOFO} (Reverse FOcus FOcus) lattice proposed for initial 6D cooling in \texttt{A complete  cooling scenario for a Muon Collider}~\cite{pac} has been studied and simulated by a number of programs~\cite{rfofo}. In this reference, 201~MHz rf is employed in \texttt{pill-box} cavities. Such cavities  will be tested by MuCool~\cite{break}, and used in  MICE~\cite{mice} - the first demonstration of ionization cooling. The advantage of such cavities is that the average acceleration gradient is close to the maximum surface fields in the cavity. 
\begin{figure}[htb]
\includegraphics*[width=75mm]{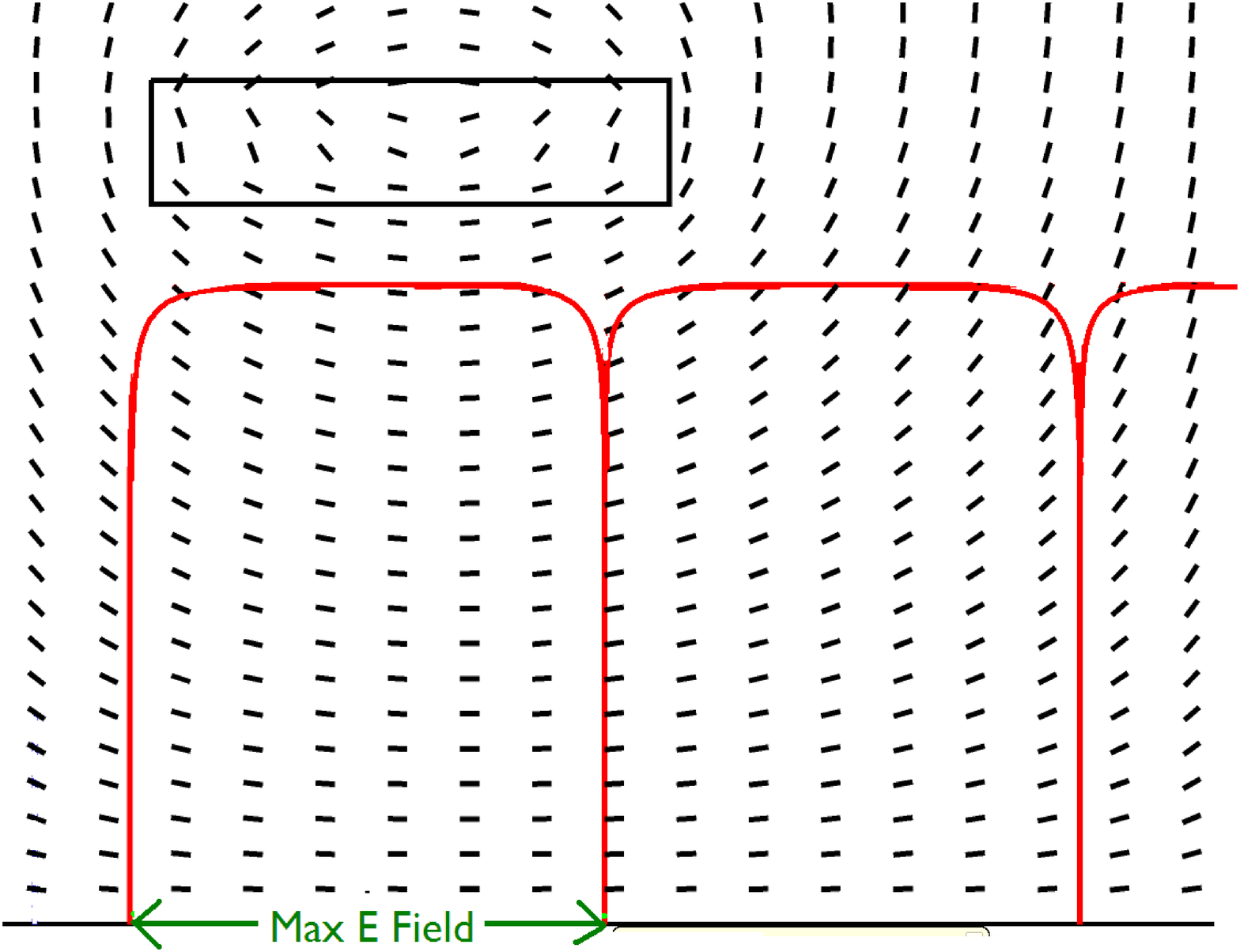}
\caption{(Color) Magnetic field directions (black) and maximum electric fields (green) for Pill-box cavities in RFOFO lattice with coils outside rf.}
\label{breaks1}
\end{figure}

The magnetic fields in the RFOFO cooling lattice are supplied by coils located outside the rf, as shown in figure~\ref{rfofo-cell}. Note that in these lattices the coils are  slightly tilted to generate bending and the dispersion needed for emittance exchange and 6D cooling. Figure~\ref{breaks1} shows the magnetic fields in this lattice, together with the approximate location of the maximum electric fields. 
It is seen that the maximum electric fields, along the axis, are aligned with the magnetic fields. It is suggested that it is this alignment that causes their poor breakdown performance observed by the MuCool collaboration\cite{break} and shown in figure~\ref{break}.
\begin{figure}[htb!]
\centering
\includegraphics*[width=75mm]{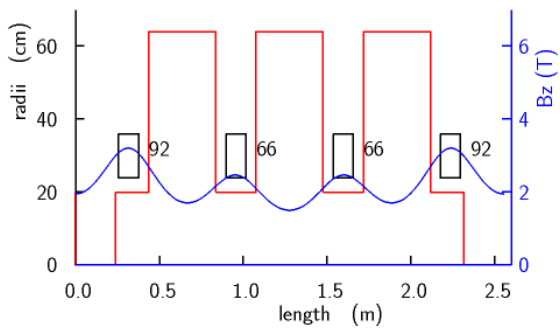}
\caption{(Color) Schematic of one full cell showing locations for absorbers.}
\label{fullcell}
\end{figure}

\begin{figure}[htb]
\includegraphics*[width=75mm]{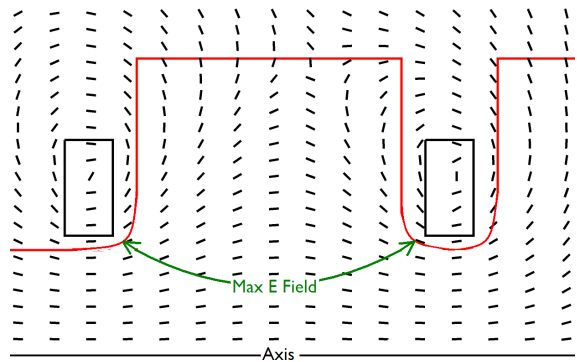}
\caption{(Color) Magnetic field directions (black) and maximum electric fields (green) for  Open cavities in RFOFO lattice with coils in the irises.}
\label{breaks2}
\end{figure}

The proposed modifications to this lattice are a) to use open iris standing wave cavities, and b) place the solenoid coils in those irises.
Figure~\ref{fullcell} shows a full cell of the modified lattice, including the spaces at length=0, and length=2.6~m where the required hydrogen wedge absorbers can be located.
Figure~\ref{breaks2} shows the magnetic fields and approximate location of maximum electric fields; it is seen that the magnetic fields are almost perpendicular to the maximum electric field, which, it is believed, will help to suppress breakdown by magnetic insulation~\cite{mag-ins1}: not completely~\cite{mag-ins2}, but some improvement is expected.  
\section{Experiments}
\begin{figure}[htb!]
\includegraphics*[height=.275\textheight]{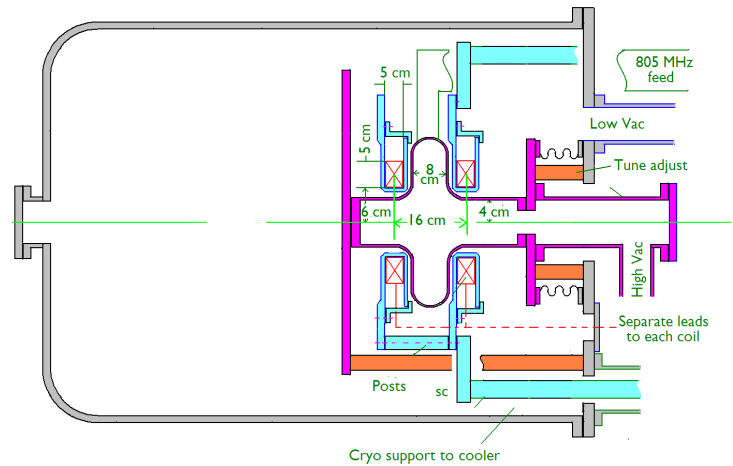} 
\caption{(Color) Experiment with one rf cavity.}
\label{exp1}
\end{figure}

\begin{figure}[hbt!]
\includegraphics*[height=.275\textheight]{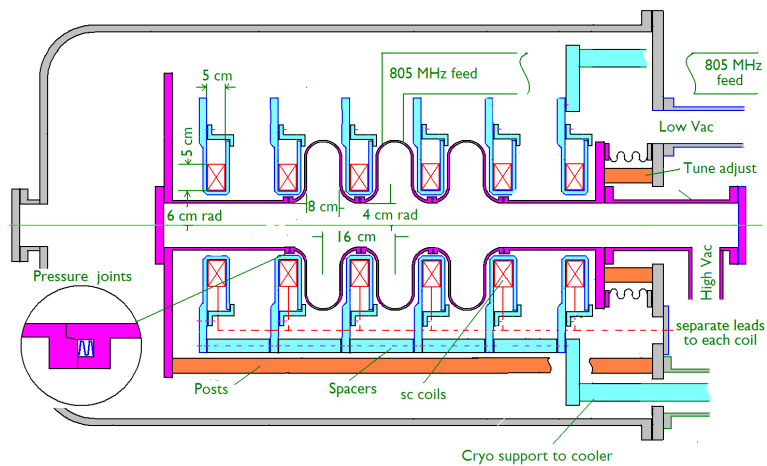} 
\caption{(Color) Experiment with full cell plus matching.}
\label{exp2}
\end{figure}

Clearly, experiments are needed to find if the suggested suppression of breakdown in these \texttt{coil in iris} geometries is real. The first such experiment, that is already planned, is to test the existing 805~MHz pill-box cavity in a field that is at right angles to its axis. The next step would be to test a single 805~MHz open cavity with two solenoid coils placed one on either side of the cavity as shown in figure \ref{exp1}. The step beyond that might be to test a full cell of the proposed lattice, including the first coils of the preceding and following lattices, as shown in figure \ref{exp2}.

\section{Summary}
It is proposed that the observed difference of rf breakdown in magnetic fields between pill-box and open iris cavities is related to their differing angles between maximum electric fields and the magnetic fields at their location. If this is correct, performance would be further improved by placing the solenoid coils in the open irises. A lattice, using this concept, for use in 6D RFOFO cooling for a Muon Collider is presented. An experimental program is proposed to test this hypothesis. 



\begin{theacknowledgments}
This work has been supported by U.S. Department of Energy under contracts AC02-98CH10886.
\end{theacknowledgments}


\begin{thebibliography}{9}

\bibitem{break}	J.~Norem et al., \textsl{Recent Results from the MuCool Test Area}, PAC07 (2007), p. 2239; \textsl{The MuCool rf Program}, EPAC 06,  (2006), p. 1358. 
\bibitem{iss}International Scoping Study (ISS), URL: http://www.cap.bnl.gov/mumu/project/ISS/ ; M.~Zisman, \textsl{International Scoping Study of a Future Accelerator Neutrino Complex}, EPAC06 (2006), p. 2427.
\bibitem{mag-ins1}F.~Winterberg, Rev. Sci. Instrum., \textbf{41}, (1970), p. 1756; 1967 Annual Report, Conf. Electrical Insulation Dielectric Phenomena, pub. 1578, National Academy of Sciences, Washington D.C. (1968).
\bibitem{mag-ins2}E.H.~Hirsch, Rev. Sci. Instrum., \textbf{42}, (1971), p. 1371.
\bibitem{pac} R.B.~Palmer et al., \textsl{A Complete Scheme of Ionization Cooling for a Muon Collider},  PAC07 (2007), p.~3193.
\bibitem{rfofo} J.S.~Berg, R.C.~Fernow, and R.B.~Palmer, \textsl{An
     Alternating Solenoid Focused Ionization Cooling Ring,} NFMCC-doc-\#239, (2002); URL: http://nfmcc-docdb.fnal.gov/cgi-bin/ShowDocument?docid=239. 
\bibitem{mice}International Muon Ionization Cooling Experiment (MICE) URL:http://hep04.phys.iit.edu/cooldemo/; M.~Zisman, \textsl{ Status of the International Muon Ionization Cooling experiment}, PAC07 (2007), p. 2996.

\end{thebibliography}
\end{document}